\documentclass[a4paper, aps, prl, showpacs, superscriptaddress, preprintnumbers, twocolumn, nofootinbib]{revtex4-2} 
\usepackage{graphicx, color}
\usepackage{amsmath, amssymb, amscd, latexsym, bm}
\usepackage[colorlinks=true,pdfstartview=FitV,linkcolor=blue,citecolor=magenta,urlcolor=blue,bookmarks=true,bookmarksnumbered=true]{hyperref}
\newcommand{\sqsNN}{\sqrt{s_{_{\;\!\!N\;\!\!\;\!\!N}}}}

\begin{document}

\preprint{RIKEN-iTHEMS-Report-24, YITP-24-112, J-PARC-TH-0308, KUNS-3017}

\title{Optimal collision-energy range for realizing macroscopic high baryon-density matter}

\author{Hidetoshi Taya}
\email{h\_taya@keio.jp}
\address{Department of Physics, Keio University, 4-1-1 Hiyoshi, Kanagawa 223-8521, Japan}
\address{RIKEN iTHEMS, RIKEN, Wako 351-0198, Japan}

\author{Asanosuke Jinno}
\email{jinno@ruby.scphys.kyoto-u.ac.jp}
\address{Department of Physics, Faculty of Science, Kyoto University, Kyoto 606-8502, Japan}

\author{Masakiyo Kitazawa}
\email{kitazawa@yukawa.kyoto-u.ac.jp}
\address{Yukawa Institute for Theoretical Physics, Kyoto University, Kyoto, 606-8502, Japan}
\address{J-PARC Branch, KEK Theory Center, Institute of Particle and Nuclear Studies, KEK, 319-1106, Japan}

\author{Yasushi Nara}
\email{nara@aiu.ac.jp}
\address{Akita International University, Yuwa, Akita-city 010-1292, Japan}

\date{\today}

\begin{abstract}
We investigate the volume and lifetime of the high baryon-density matter created in heavy-ion collisions and estimate the optimal collision-energy range to realize the high baryon-density region over a large spacetime volume.  We simulate central collisions of gold ions for the center-of-mass energy per nucleon pair $\sqsNN=2.4\,\mathchar`-\,19.6\;{\rm GeV}$ with a microscopic transport model JAM.  We discover that the optimal range is around $\sqsNN=3\,\mathchar`-\,5\;{\rm GeV}$, where a baryon density exceeding three times the normal nuclear density is realized with a substantially large spacetime volume.  Higher and lower energies are disfavored due to short lifetime and low density, respectively.  We also point out that event-by-event fluctuations of the spacetime density profile are large, indicating the importance of the event selection in the experimental analysis.  
\end{abstract}

\maketitle

{\it Introduction.}---\ 
Exploring the properties of matter at extremely high baryon densities that exceed several times the normal nuclear density $\rho_0 \approx 0.17\;{\rm fm}^{-3}$ is one of the most exciting frontiers in current physics.  Such high-density environment is realized in, for example, the core of neutron stars and the core-collapse supernovae.  Recent remarkable progress in the observation of neutron stars through gravitational waves from binary neutron-star mergers~\cite{LIGOScientific:2017vwq,*LIGOScientific:2017ync,*LIGOScientific:2018cki,*LIGOScientific:2018hze} and pulsar X-rays~\cite{Miller:2019cac,*Miller:2021qha,*Riley:2019yda,*Riley:2021pdl} is providing valuable information for constraining the equation of state of the dense matter~\cite{MUSES:2023hyz}.  The form of matter in such extreme conditions is determined by quantum chromodynamics (QCD) --- the theory of strong interaction --- and is expected to have rich phase structure~\cite{Fukushima:2010bq}.  Unveiling their properties is one of the ultimate goals of nuclear physics.  Experimental inputs are crucial to achieve this, since the first-principle lattice-QCD calculation is difficult due to the notorious sign problem~\cite{Nagata:2021ugx} and the low-energy effective models, such as the chiral effective theory, become invalid when the baryon density exceeds about $2\rho_0$~\cite{Tews:2018kmu,Drischler:2021kxf,Koehn:2024set}.

Relativistic heavy-ion collisions are the only terrestrial experiments to create such high baryon-density matter and to study their properties in laboratories~\cite{luo2022properties}.  In these experiments, dense and hot matter is created by colliding heavy nuclei accelerated up to almost the speed of light with a large accelerator.  The experiments have been carried out at various facilities such as the LHC~\cite{LHC}, RHIC~\cite{Chen:2024zwk}, and SIS18~\cite{HADES:2019auv}, for a wide range of collision energies with the center-of-mass energy per nucleon pair $\sqsNN$ up to $5\;{\rm TeV}$.  In the past decades, the high energy region $\sqsNN \gtrsim 100\;{\rm GeV}$ has been extensively investigated at the LHC and RHIC.  They are suitable for creating extremely hot but low baryon-density matter, and have succeeded in reproducing the primordial form of matter in the early Universe --- the quark-gluon plasma~\cite{Yagi:2005yb}.  Recently, the lower energy range $\sqsNN\approx3\,\mathchar`-\,20\;{\rm GeV}$, which we call {\it intermediate collision energies}, is attracting renewed attention.  The intermediate energies are expected to be suitable for exploring high baryon-density and low-temperature matter.  This energy range is actively investigated in the Beam-Energy Scan (BES) program at RHIC~\cite{Aparin:2023fml} and NA61/SHINE at SPS~\cite{NA61}, and will be further studied worldwide~\cite{Galatyuk:2019lcf} in NICA~\cite{NICA}, FAIR~\cite{FAIR}, HIAF~\cite{HIAF}, and J-PARC-HI~\cite{J-Parc-HI}.

Several experimental results support the formation of high baryon-density matter at the intermediate energy.  First, the rapidity distribution of net-proton number in collisions of gold (Au) nuclei at $\sqsNN=5\,\mathchar`-\,200\;{\rm GeV}$~\cite{BRAHMS:2003wwg} implies strong stopping of baryon charges at mid-rapidity for the AGS energy $\sqsNN\approx5\;{\rm GeV}$ and hence realization of high baryon density.  Second, the thermal-statistical fit to the hadron yields provides a more quantitative estimate, showing that the baryon density at the chemical freezeout is maximized at $\sqsNN\approx6\,\mathchar`-\,10\;{\rm GeV}$~\cite{Randrup:2006nr,Randrup:2009ch,Begun:2012rf,Blaschke:2024jqd}.  However, the baryon density obtained in these analyses is {\it not} the maximum for a whole heavy-ion reaction, since the chemical freezeout occurs at a late stage, where the baryon density is diluted by the expansion. 

To investigate the maximum baryon density, one must resort to dynamical models of heavy-ion collisions, such as microscopic transport models~\cite{Li:1995pra, Bass:1998ca,Buss:2011mx,SMASH:2016zqf,Aichelin:2019tnk}, hydrodynamic models~\cite{Ivanov:2005yw,Ivanov:2013wha,Ivanov:2013yqa,Ivanov:2013yla}, and hybrid models~\cite{Petersen:2008dd,Batyuk:2016qmb,Akamatsu:2018olk,Cimerman:2023hjw}, as well as semi-analytical models~\cite{Mendenhall:2021maf}.  Previous microscopic transport-model calculations predicted that the maximum baryon density {\it at the center} of the collision system can exceed $8\rho_0$ for $\sqsNN\gtrsim5\;{\rm GeV}$~\cite{Li:1995pra,Danielewicz:1998vz,Arsene:2006vf,Ohnishi:2015fhj,Bhaduri:2022cql}.  It has also been discussed that even higher densities over $10\rho_0$ can be reached in some collision events due to large event-by-event fluctuations~\cite{Ohnishi:2015fhj}.  However, these studies focus on the density at the center, and {\it it is unclear how much the dense region extends in space and time}.  The large spatial and temporal sizes, besides the maximum baryon density, are crucial for the dense matter to leave observable signals in the final state.  Despite its importance, to the best of our knowledge, quantitative studies on the volume and lifetime of the dense region and their dependence on $\sqsNN$ have not been performed~\footnote{As a potentially relevant study, we note that Ref.~\cite{Friman:1997sv} investigated the spacetime volume of the quark condensate in the low-density regime.  }.  

The purpose of this Letter is to examine the spacetime volume of the high baryon-density region in intermediate-energy heavy-ion collisions.  We will discover that the optimal energy range that maximizes the spacetime volume is around $\sqsNN=3\,\mathchar`-\,5\;{\rm GeV}$, for which the typical value of the density is about a few times the normal nuclear density.  These range and density are significantly lower than the previous estimates~\cite{Randrup:2006nr,Randrup:2009ch,Begun:2012rf,Blaschke:2024jqd, Li:1995pra,Danielewicz:1998vz,Arsene:2006vf,Ohnishi:2015fhj,Bhaduri:2022cql}, and hence would provide crucial information for the dense-matter study in heavy-ion collisions, e.g., detector design.

{\it Measures.}---
To quantify the volume and lifetime of the high-baryon-density region in heavy-ion collisions, we propose the following three measures.  These measures are constructed from the baryon density in the local rest frame (Eckart frame)~\footnote{Equation~\eqref{eq:rho} is ill-defined for $J^2<0$.  We have carefully checked that $J^2<0$ occurs only when anti-baryons are produced and is not caused by numerical errors.  Anti-baryon production is well suppressed for $\sqsNN<20\;{\rm GeV}$.  Hence, it is rare to have $J^2<0$.  Even if it happens, the value is found to be very close to zero within the numerical accuracy.  Therefore, in our simulation, we simply set $\rho(x)=0$ if we have $J^2(x)<0$.  }, 
\begin{align}
    \rho(x) := \frac{J^0(x)}{\gamma[J(x)]} \;, \label{eq:rho}
\end{align}
where $J^\mu(x)$ is the net baryon current at each spacetime point $x$ in the center-of-mass frame of a collision system and $\gamma[J] := 1/\sqrt{1- ({\bm J} /J^0)^2 }$ is the Lorentz factor. 

The first measure is the spatial volume that exceeds a threshold density value $\rho_{\rm th}$,
\begin{align}
	&V_3(\rho_{\rm th};t) := \int_{\rho(x)>\rho_{\rm th}} {\rm d}^3{\bm x} \, \gamma[J(x)] \;. \label{eq:V3}
\end{align}
where $\gamma[J]$ is inserted to define the volume in the local rest frame.  Second, we introduce a Lorentz invariant four-volume $V_4(\rho_{\rm th})$ as~\footnote{The integral in Eq.~\eqref{eq:V4} diverges for $\rho_{\rm th}\lesssim\rho_0$, since nuclear matter has the density $\rho_0$ for $t<0$ and also nuclear clusters can be formed during heavy-ion reactions, which can survive for a long time.  We, thus, limit ourselves to $\rho_{\rm th} > 2\rho_0$, where these effects are negligible.}
\begin{align}
	V_4(\rho_{\rm th}) := \int_{\rho(x)>\rho_{\rm th}} {\rm d}t{\rm d}^3\bm{x} \;. \label{eq:V4}
\end{align}
Finally, we define a measure of the typical lifetime of the dense region as
\begin{align}
	\tau(\rho_{\rm th}) := \frac{V_4(\rho_{\rm th})}{\max[ V_3(\rho_{\rm th};t)]} \;, \label{eq:tau}
\end{align}
where $\max[V_3(\rho_{\rm th};t)]$ is the maximum of $V_3(\rho_{\rm th};t)$ over time.  We compute Eqs.~\eqref{eq:V3}--\eqref{eq:tau} on an event-by-event basis, i.e., they are constructed from $J^\mu(x)$ for each event. 

Advantages of the measures are their simplicity and the clarity of the physical meanings.  They can be calculated straightforwardly in a given dynamical model without any further assumptions, once $J^\mu(x)$ is given.  However, these measures are not concerned with the local equilibration, and hence the local temperature nor the chemical potential.  Whereas local equilibration has been studied within some microscopic transport models~\cite{Sorge:1995pw,Bravina:1998pi,Bravina:1999dh,Bravina:2000iw,Bravina:2008ra,Nara:2017qcg}, it must introduce criteria for equilibration in a finite and dynamical system, leaving ambiguities in the analysis and interpretation.  Rather, we here avoid these complications to present unambiguous results by ignoring equilibration.  Nevertheless, these measures can still constrain the spacetime profile of the {\it equilibrated} dense region, since $V_3(\rho_{\rm th},t)$ and $V_4(\rho_{\rm th})$ obviously give the upper limits for the spacetime volumes of the equilibrated dense region.  Furthermore, the formation of the dense region is of interest even without equilibration because such high density can induce intriguing phenomena, such as the chiral vortical effect~\cite{Kharzeev:2015znc} and the nucleation of deconfined regions.

{\it Simulation method.}---\
We investigate the $\sqsNN$ dependence of the measures~\eqref{eq:V3}--\eqref{eq:tau} in central Au\;\!+\;\!Au collisions.  We employ a microscopic transport model JAM~\cite{Nara:1999dz,JAM} and generate one thousand collision events with the impact parameter less than $3\;{\rm fm}$, which roughly corresponds to the top 5\% centrality cut.  The simulations are performed with the default setting of the version 2.5743, unless otherwise stated.  This version incorporates the recently-implemented covariant cascade algorithm~\cite{Nara:2023vrq} and the Lorentz-vector version of the relativistic molecular dynamics (RQMDv) with the mean field of a soft momentum-dependent potential~\cite{Nara:2021fuu,Nara:2022kbb}.  As shown in Refs.~\cite{Nara:2021fuu,Nara:2022kbb}, RQMDv can successfully reproduce various experimental data, such as the collision energy dependencies of the directed flows of protons and $\Lambda$ over a wide range of collision energies~\footnote{Nonetheless, we remind that the first-principle calculation of the equation of state of dense QCD matter is unavailable and cannot reject the possibility that other modellings or parametrizations would be able to fit the data.  Therefore, we need to keep in mind that our results would still contain some uncertainties, which we left for future work to be removed.  }.  

To define the baryon current $J^\mu(x)$ as a continuous quantity, we smear the baryon density of particles at position $X_b$ and momentum $P^\mu_b$ as $J^\mu(x) := \sum_{b} V^\mu_b B_b g(x-X_b;P_b)$, where $B_b$ and $V^\mu_b=P^\mu_b/P_b^0$ are the baryon number and velocity of the $b$-th particle~\footnote{In general, the velocity $V^\mu$ can receive correction from the mean field~\cite{Bertsch:1988ik,Weber:1992qc}.  However, we neglect it since we have checked that its effect is well suppressed.}, respectively, and $g(x;P) := \gamma[P]/(\sqrt{2\pi}r)^3 \exp[- (|\bm{x}|^2 + (\gamma[P] \bm{V}_b\cdot\bm{x})^2)/(2r^2) ]$ is a relativistic Gaussian smearing function~\cite{Fuchs:1995fa, Oliinychenko:2015lva}.  We use the smearing width $r = 1\;{\rm fm}$ for all particles, which is almost the same size as the charge radius of protons~\cite{Gao:2021sml}.  The smearing can be interpreted as intrinsic distribution of the baryon density inside a particle or a technical coarse-graining for numerics.  The maximum value of the smearing function $g(0;0)\approx0.37\rho_0$ is much smaller than $\rho_0$, indicating that a considerable number of baryons must overlap to realize $\rho(x)\gg\rho_0$.

\begin{figure}[t]
\begin{flushleft}
\includegraphics[width=0.2697\textwidth, clip, trim = 0  44  0  0]{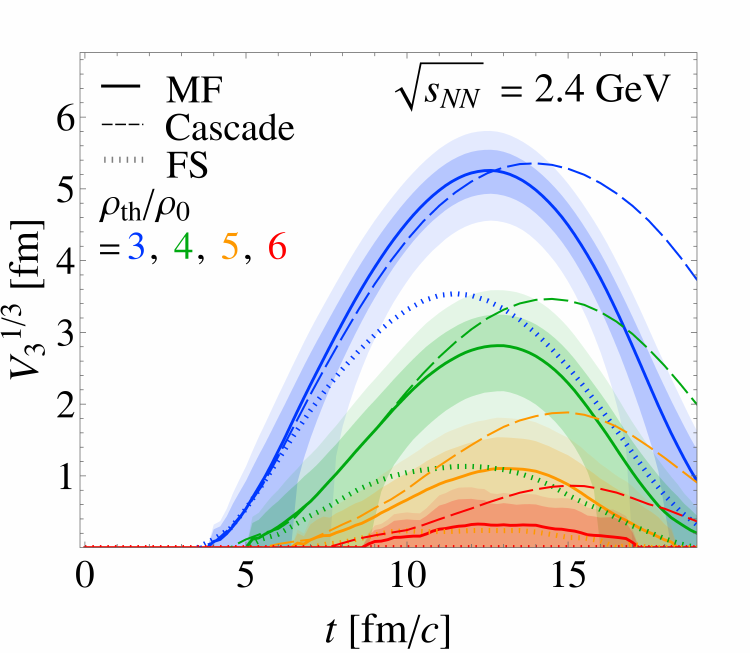} \hspace*{-5.3mm}
\includegraphics[width=0.2300\textwidth, clip, trim = 37 44 16  0]{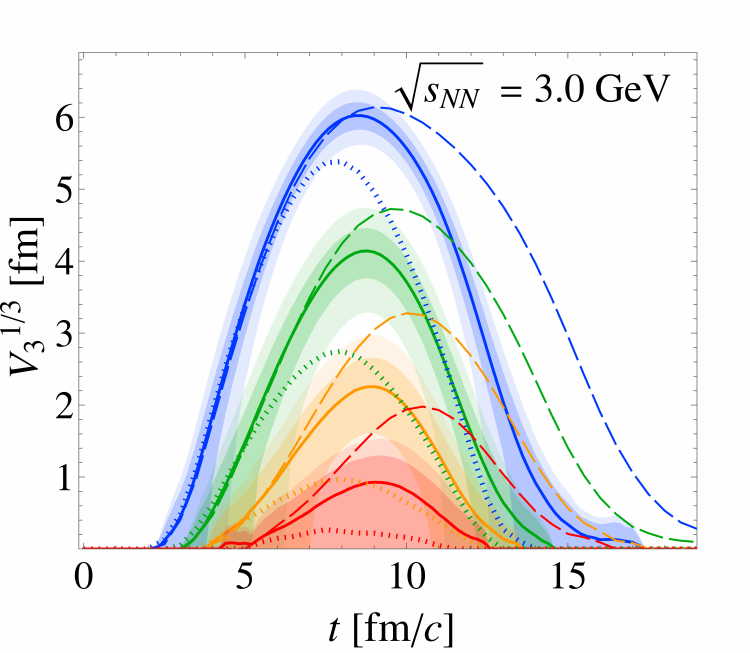} \vspace*{-4.7mm} \\
\includegraphics[width=0.2697\textwidth, clip, trim = 0   0  0 20]{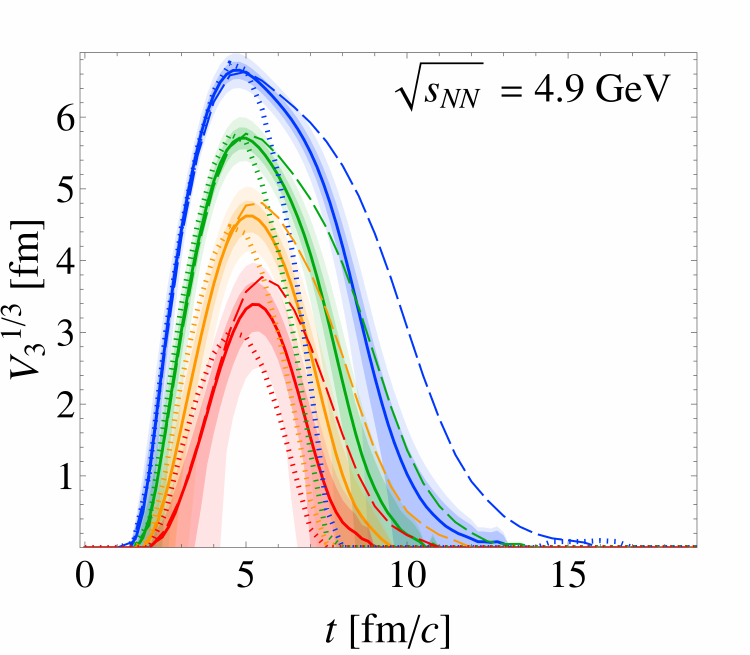} \hspace*{-5.3mm}
\includegraphics[width=0.2300\textwidth, clip, trim = 37  0 16 20]{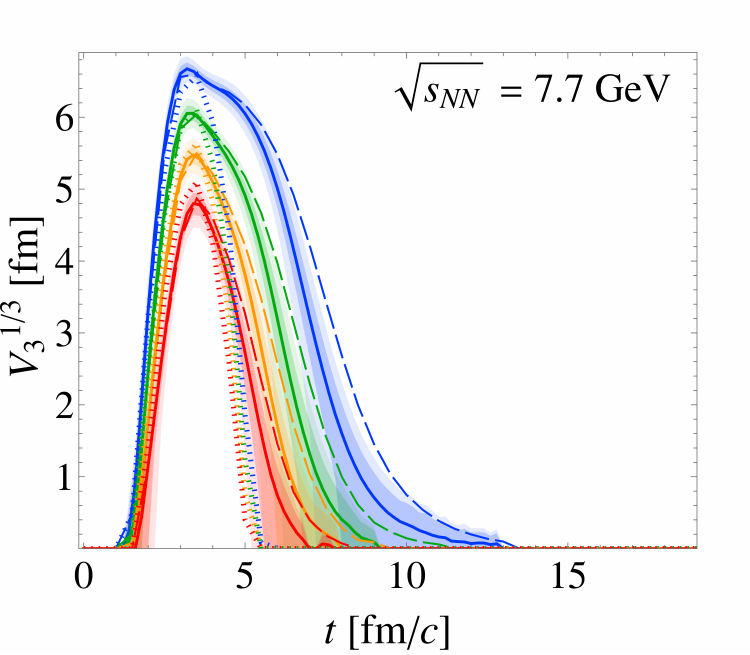}
\end{flushleft}
\vspace*{-4.7mm} 
\caption{Time evolution of the spatial volume $V_3(\rho_{\rm th};t)$ in central Au\;\!+\;\!Au collisions for several thresholds $\rho_{\rm th}$ and collision energies $\sqsNN$.  The thick-solid, dashed, and dotted lines represent MF, Cascade, and FS results, respectively.  The color bands around the MF results show the $1\sigma$ (denser) and $2\sigma$ (lighter) bands of event-by-event fluctuations.  At the initial time $t=0$, the centers of the colliding nuclei are separated by $2R_{\rm Au}/\gamma_{\rm Au}+3\;{\rm fm}$ in the longitudinal direction with $\gamma_{\rm Au}$ being the Lorentz factor for the incident Au ions.}
\label{fig:V3}
\end{figure}

{\it Simulation results.}---\
Figure~\ref{fig:V3} shows the time dependencies of $V_3(\rho_{\rm th};t)$ for various collision energies and threshold densities.  The thick-solid, thin-dashed, and dotted lines show, respectively, the results from the full simulation with the mean field (MF), the cascade mode (Cascade), where the mean field is turned off, and the free-streaming mode (FS), where all the interactions are switched off.  Cascade and FS are shown as guidelines to clarify the physics behind the full simulation result (MF).  In particular, the deviations from the FS result, which only takes into account a simple overlapping of two Lorentz-contracted nuclei, indicate the effects of the baryon stopping induced by interactions.  

One observes that $V_3(\rho_{\rm th};t)$ for $\rho_{\rm th}/\rho_0=3$ has the maximum value ${\rm max}[V_3(\rho_{\rm th};t)]^{1/3}\approx6\;{\rm fm}$ for all collision energies.  We regard this length scale as ``macroscopically" large~\footnote{By macroscopic, we only mean the largeness of the length scale and do not require that the number of particles consisting the matter is large, i.e., we do not mean the thermodynamic limit.  }, since it is much larger than the typical microscopic length scale of the system $\approx 1\;{\rm fm}$, determined by the interaction range and the sizes of hadrons.  Additionally, the achieved length scale is comparable to the radius of the colliding gold ions $R_{\rm Au}\approx6.4\;{\rm fm}$ and thus is ``maximally" large in the sense that the overlap of two nuclei cannot result in a volume larger than $\approx R_{\rm Au}^3$.  Regions with $\rho_{\rm th}/\rho_0>3$ are created with smaller $V_3(\rho_{\rm th};t)$, which increase with $\sqsNN$.  We also find that the dense region disappears more quickly for larger $\sqsNN$.  

When comparing MF with Cascade, one finds that $V_3(\rho_{\rm th};t)$ is more suppressed in MF.  This is because the repulsive mean field~\cite{Nara:2021fuu} makes the matter less compressed~\cite{fullpaper}.  The mean-field effect becomes more important for lower collision energies, especially for $\sqsNN \lesssim 3.0\;{\rm GeV}$.  The comparison between MF and FS shows that $V_3(\rho_{\rm th};t)$ is enhanced by the interaction.  Due to collisions, the particles are decelerated and gather around the collision point, forming up a dense region.  Note that the achievable density for FS can na\"ively be estimated as $\rho/\rho_0 \approx \sqsNN/m_N$ (with $m_N \approx 1\;{\rm GeV}$ being the nucleon mass) due to the Lorentz contraction, provided that the colliding nuclei are uniform.  In reality, however, the maximum density can exceed this value significantly (i.e., $V_3 > 0$ even for $\rho_{\rm th}/\rho_0 > \sqsNN/m_N$) because of the non-uniform density distribution inside the colliding ions due to event-by-event fluctuations.    

\begin{figure}[t]
\centering
\includegraphics[width=0.40\textwidth]{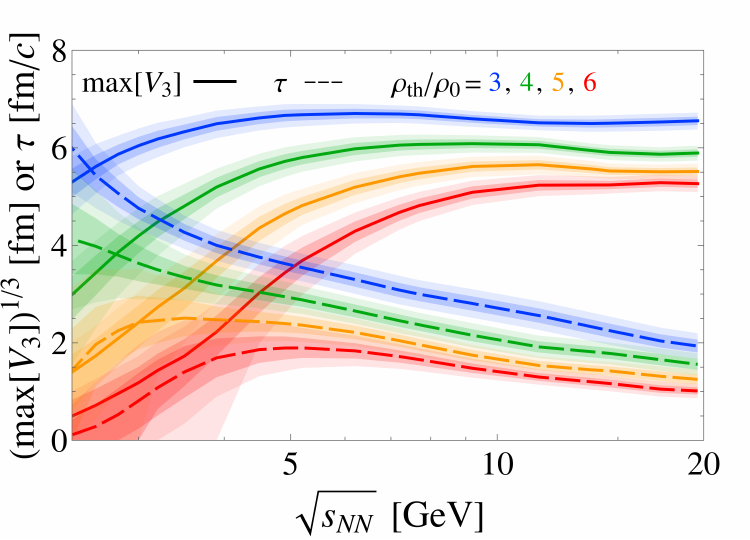}
\caption{MF results for the collision energy $\sqsNN$ dependence of the maximum volume ${\rm max}[V_3]$ (solid) and the lifetime $\tau$ (dashed) in central Au\;\!+\;\!Au collisions.  The bands show the event-by-event fluctuations as in Fig.~\ref{fig:V3}.  }
\label{fig:V3tau}
\end{figure}

\begin{figure}[t]
\centering
\includegraphics[width=0.45\textwidth]{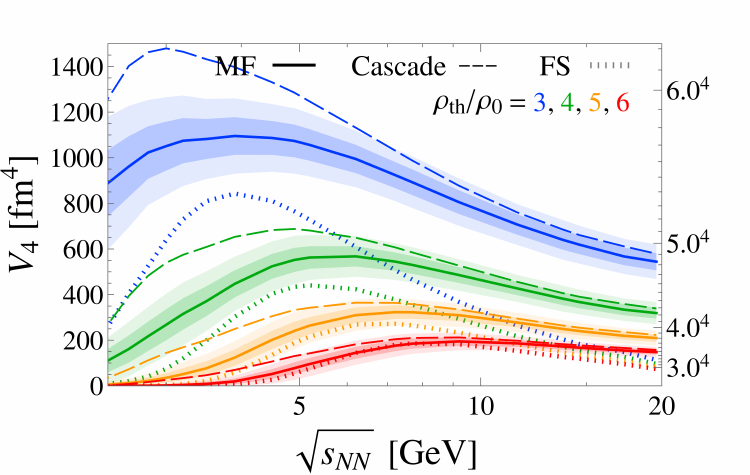}
\caption{The collision energy $\sqsNN$ dependence of the spacetime volume $V_4(\rho_{\rm th})$ in central Au\;\!+\;\!Au collisions.  The meanings of the lines and bands are the same as Fig.~\ref{fig:V3}.}
\label{fig:V4}
\end{figure}

In order to examine the $\sqsNN$ dependence observed in Fig.~\ref{fig:V3} quantitatively, we show ${\rm max}[V_3(\rho_{\rm th};t)]$ and $\tau(\rho_{\rm th})$ as functions of $\sqsNN$ in Fig.~\ref{fig:V3tau}.  It shows that ${\rm max}[V_3(\rho_{\rm th};t)]$ grows and saturates with increasing $\sqsNN$.  Meanwhile, $\tau(\rho_{\rm th})$ decreases except for $\rho_{\rm th}/\rho_0\ge5$ at low collision energies, where the formation of the region with $\rho(x)>\rho_{\rm th}$ scarcely occurs.

To obtain an observable signal of dense matter, it is reasonable to expect that both the spatial volume and the lifetime are large~\footnote{To be precise, the signals become larger by definition, for {\it extensive} quantities, such as particle yields, as we increase the spacetime volume of the fireball.  On the other hand, {\it intensive} quantities such as conductivity should not be affected by the spacetime volume.  Nonetheless, it is impossible to directly measure such intensive quantities in heavy-ion-collision experiments; instead, what we can actually measure is the particle yields, which are extensive and with which we extract the information of intensive quantities indirectly.  Having said that the largeness of the spacetime volume is a necessary condition to have a sizable signal for extensive quantities, we note that there arise contaminations to the signal from various other factors such as the lifetime of the hadronic stage, making the signal-to-noise ratio worse.  It is important to consider these factors when making a quantitative prediction.  }.  The optimal $\sqsNN$ to maximize ${\rm max}[V_3(\rho_{\rm th};t)]$ and $\tau(\rho_{\rm th})$ simultaneously may be estimated from the crossing point of the solid- and dashed-blue lines in Fig.~\ref{fig:V3tau}.  Taking $\rho_{\rm th}/\rho_0=3$ as an example, it is located at $\sqsNN\approx2.6\;{\rm GeV}$, where a macroscopically large spacetime volume, ${\rm max}[V_3(\rho_{\rm th};t)]^{1/3}=\tau(\rho_{\rm th})\approx5.5\;{\rm fm}$, is realized.  For $\rho_{\rm th}/\rho_0=4$, the crossing point is located at $\sqsNN\approx2.8\;{\rm GeV}$ with ${\rm max}[V_3(\rho_{\rm th};t)]^{1/3}=\tau\approx4\;{\rm fm}$, which is smaller than the result for $\rho_{\rm th}/\rho_0=3$, but still occupies a certain spacetime volume of the reaction zone.  Similar arguments suggest that an experimental analysis of the matter with $\rho(x)/\rho_0\gtrsim5$ is challenging, as ${\rm max}[V_3(\rho_{\rm th};t)]$ and $\tau(\rho_{\rm th})$ are suppressed and are no longer macroscopically large.

One can make a similar estimate of the optimal $\sqsNN$ for $V_4(\rho_{\rm th})$.  Figure~\ref{fig:V4} shows $\sqsNN$ dependence of $V_4(\rho_{\rm th})$, which indicates that the MF result for each $\rho_{\rm th}$ has a plateau around the maximum.  We identify the optimal energy {\it range} for $V_4(\rho_{\rm th})$ with these plateaus and find that the optimal ranges are given by $\sqsNN \approx 3\,\mathchar`-\,5\;{\rm GeV}$ and $\sqsNN \approx 5\,\mathchar`-\,7\;{\rm GeV}$ for $\rho_{\rm th}/\rho_0=3$ and $4$, respectively.  In the optimal energy ranges, the four volumes are $[V_4(\rho_{\rm th})]^{1/4}\approx5.8\;{\rm fm}$ and $4.9\;{\rm fm}$, respectively, which are almost the same as the values of $[V_3(\rho_{\rm th},t)]^{1/3}$ and $\tau(\rho_{\rm th})$ at their optimal energy estimated by the crossing points in Fig.~\ref{fig:V3tau}.  The optimal range estimated from $V_4(\rho_{\rm th})$ for each $\rho_{\rm th}$ does not include the optimal energy estimated from ${\rm max}[V_3(\rho_{\rm th};t)]$ and $\tau(\rho_{\rm th})$, but is slightly deviated to higher $\sqsNN$.  This is not an inconsistent result, since the ``optimality" depends on quantities to focus on in general.  Nevertheless, we note that ${\rm max}[V_3(\rho_{\rm th};t)]$ and $\tau(\rho_{\rm th})$ remain considerably large in the optimal energy range of $V_4(\rho_{\rm th})$~\footnote{As a logical possibility, it is possible that either of $V_3$ or $\tau$ is small and that the smallness of it is compensated by the other to have a large $V_4$.  Our result eliminates this possibility.  }.  

Wrapping up these results, we conclude that the region with $\rho(x)/\rho_0\ge3$ can be formed with macroscopically large spatial and temporal sizes in central Au\;\!+\;\!Au collisions.  The optimal range of $\sqsNN$ to maximize $V_4(\rho_{\rm th})$ starts from $\sqsNN\approx3\;{\rm GeV}$ and extends up to $\sqsNN\approx5\;{\rm GeV}$, where ${\rm max}[V_3(\rho_{\rm th};t)]$ and $\tau(\rho_{\rm th})$ also take macroscopically large values~\footnote{One could further narrow the range or pinpoint a specific optimal energy value by requiring additional conditions/assumptions (e.g., specify what physics process to focus, consider equilibration, and take into account other characteristics such as vorticity~\cite{Deng:2020ygd} and electromagnetic field~\cite{Taya:2024wrm, Taya:2025utb}).  However, it inevitably involves detailed case-by-case analyses, and hence is omitted for the purpose of the present Letter.  Nevertheless, we stress that the optimal range has already been constrained strongly by the spacetime volume, and such additional analyses would not alter the main conclusion of the present work.  }.  The optimal collision-energy range for $\rho(x)/\rho_0\ge4$ is estimated to be slightly higher than that for $\rho(x)/\rho_0\ge3$ and is located at $\sqsNN \approx 5\,\mathchar`-\,7\;{\rm GeV}$.  Compared to the previous research, the estimated optimal $\sqsNN$ is considerably lower~\cite{Randrup:2006nr,Ohnishi:2015fhj}.  Although the maximum baryon density can locally exceed $\rho(x)/\rho_0=8$, as shown in Refs.~\cite{Danielewicz:1998vz,Arsene:2006vf,Ohnishi:2015fhj,Bhaduri:2022cql}, the volume and lifetime of such high baryon-density regions are small.    

Finally, we discuss event-by-event fluctuations, which are driven by the positional fluctuation of nucleons on the colliding nuclei~\cite{fullpaper}.  In Figs.~\ref{fig:V3}--\ref{fig:V4}, the colored bands show the $1\sigma$ (dense) and $2\sigma$ (light) ranges of the event-by-event distribution of the individual quantities.  The figures show that the locations of the optimal energy ranges stay robust against the fluctuations, while the values of $V_3(\rho_{\rm th};t)$, $\tau$, and $V_4(\rho_{\rm th})$ can fluctuate significantly.  For example, the upper and lower bounds of the $2\sigma$ band of $V_4(\rho_{\rm th})$ at $\sqsNN\lesssim3\;{\rm GeV}$ and $\rho_{\rm th}/\rho_0=3$ are $1250$ and $850\;{\rm fm}^4$, respectively.  In other words, the values of $V_4(3\rho_0)$ in the top and bottom $2.5\%$ collision events differ by a factor of more than $1250/850\approx1.5$.  This means that, if a selection of such large- and small-$V_4$ events is available, their comparison can be used to investigate the properties of dense matter from the spacetime-volume dependence of experimental observables.  As another example, for $\rho_{\rm th}/\rho_0=5$, the lower $2\sigma$-bound of ${\rm max}[V_3(5\rho_0,t)]$ vanishes at $\sqsNN\approx3\;{\rm GeV}$, whereas the upper bound is about $(3\;{\rm fm})^3$.  Exploiting these large event-by-event fluctuations may allow us to explore even denser regions in the experiments.  

{\it Discussions.}---\ 
We have shown that a high-density matter with $\rho(x)/\rho_0\gtrsim3$ can be created with macroscopically large volume at intermediate collision energies.  Since $\rho(x)/\rho_0\gtrsim3$ is the density expected to be realized in the core of neutron stars~\cite{Baym:2017whm,MUSES:2023hyz}, it is an important next step to pursue the connection between the two realms.  In heavy-ion collisions, observables, such as the directed flow~\cite{Sahu:1999mq, Nara:2016hbg}, dilepton yields~\cite{Nishimura:2022mku,Savchuk:2022aev,Nishimura:2023oqn,Nishimura:2023not}, and fluctuations of conserved charges~\cite{Asakawa:2015ybt,Luo:2017faz,Bluhm:2020mpc}, are expected to be sensitive to the equations of state and the phase transitions.  Examining these observables based on our results on the spacetime density profile enables us to further constrain the properties of the matter at baryon densities relevant to neutron stars.  In the present study, we have focused on the measures~\eqref{eq:V3}--\eqref{eq:tau}, which do not care about the equilibration of matter.  Although this ignorance is the key to realizing the quantitative analysis of the spacetime profile, the consideration of equilibration, as well as the estimate of the resulting temperature and chemical potential, is another important extension of this study to approach the matter properties in equilibrium~\cite{fullpaper}. 

Let us discuss the model dependence of our results that rely on JAM with a particular mean field, i.e., RQMDv with a soft momentum-dependent potential.  We have executed JAM with other mean-field parameters, such as a hard momentum-dependent potential and momentum-independent potentials~\cite{Nara:2021fuu} for comparison.  We have found that the choice of the mean field only gives minor differences for $\sqsNN>2.4\;{\rm GeV}$, which are smaller than that between the MF and Cascade results in Figs.~\ref{fig:V3} and~\ref{fig:V4}~\cite{fullpaper}.  Therefore, the dependence on the mean field is expected to be well suppressed.  Nevertheless, since these results are obtained within JAM, it is interesting to reinforce our findings with other transport models.  

Another characteristic of JAM is that it contains no explicit quark and gluon degrees of freedom.  Concerning this, we note that recent experimental data on elliptic flow suggest that the quark-number scaling, which is a possible signal of the formation of the quark-gluon plasma~\cite{Fries:2003kq}, seems violated for $\sqsNN \lesssim5\;{\rm GeV}$~\cite{STAR:2013ayu,STAR:2015rxv,STAR:2020dav,STAR:2021yiu}.  Therefore, around the optimal range $\sqsNN\approx3\,\mathchar`-\,5\;{\rm GeV}$ the quark-gluon plasma may be unformed, or formed only within a small fraction of the total system.  Such a picture is also supported for $\sqsNN\lesssim10\;{\rm GeV}$ from the transport-model calculations having both partonic and hadronic degrees of freedom~\cite{Konchakovski:2012yg,Aichelin:2019tnk}.  In this case, the hadronic treatment would suffice.  The analysis in Ref.~\cite{Nara:2017qcg} also suggests that the impact of the phase transition on the spacetime evolution of the matter by softening the equations of state is small.  Nevertheless, it is worthwhile to test those expectations using models that include explicit quark degrees of freedom such as  AMPT~\cite{Lin:2004en,Wang:2021owa}, PHSD~\cite{Cassing:2008sv,Cassing:2009vt,Konchakovski:2012yg}, PHQMD~\cite{Aichelin:2019tnk}, PACIAE~\cite{Lei:2023srp}, and hybrid models~\cite{Steinheimer:2011mp,Akamatsu:2018olk,Schafer:2021csj,Ege:2024vls,Cimerman:2023hjw}, and that incorporate the phase transition dynamics such as UrQMD~\cite{Savchuk:2022aev,Savchuk:2022msa,Li:2022iil,Steinheimer:2022gqb} and JAM with an advanced option~\cite{Nara:2017qcg}.  

We comment on the relation of this study with the event-by-event analysis of fluctuations in conserved charges~\cite{Asakawa:2015ybt,Luo:2017faz,Bluhm:2020mpc}.  The large fluctuations of the spacetime volume shown in Figs.~\ref{fig:V3}--\ref{fig:V4} may fake signals of the fluctuation observables\footnote{We note that a heavy-ion system is genuinely nonequilibrium, especially for intermediate energies.  In such nonequilibrium situations, conserved quantities are not necessarily determined only by the thermal parameters and finite-volume effects cannot be simply eliminated by taking the ratio of the moments.  }, which are important for the search for the QCD critical point.  Taming the fluctuations of the spacetime volume, which would in part be regarded as the volume fluctuations~\cite{Skokov:2012ds,Braun-Munzinger:2016yjz}, may be crucial in this energy range.

{\it Summary.}---\ 
We have investigated the volume and lifetime of the high baryon-density matter created in central Au\;\!+\;\!Au collisions by introducing the three measures~\eqref{eq:V3}--\eqref{eq:tau} and estimated the optimal collision-energy range to realize the largest spacetime volume using a microscopic transport model JAM.  The numerical results suggest that the dense region with $\rho(x)/\rho_0\gtrsim3$ can be created with large volume and lifetime with the optimal collision energy around $\sqsNN\approx3\, \mathchar`-\,5\;{\rm GeV}$.  Since this range is significantly lower than the previous estimates~\cite{Randrup:2006nr,Ohnishi:2015fhj}, these findings are crucial for the designs and interpretations of the near-future experiments such as NICA, FAIR, HIAF, and J-PARC-HI, as well as the current RHIC-BES program and NA61/SHINE.  

{\it Acknowledgments.}---\ 
The authors thank Akira~Ohnishi and Toru~Nishimura for the collaboration at the early stage of the present work.  They also thank Takao~Sakaguchi and Hiroyuki~Sako for useful discussions.  This work is supported by JSPS KAKENHI under Grants No.~22K14045 and No.~24K17058 (HT), Grants No.~JP19H05598, No.~JP22K03619, No.~JP23H04507, and No.~JP24K07049 (MK), Grant No.~JP21K03577 (YN), the RIKEN special postdoctoral researcher program (HT), JST SPRING, Grant No.~JPMJSP2110 (AJ), and the Center for Gravitational Physics and Quantum Information (CGPQI) at Yukawa Institute for Theoretical Physics (MK).  A part of the numerical simulations have been carried out on Yukawa-21 at Yukawa Institute for Theoretical Physics (YITP), Kyoto University.  

\bibliography{bib}

\end{document}